\documentclass{moriond}

\newcommand{\lcdm}{$\Lambda$CDM}

\usepackage{xspace}
\usepackage{amsfonts, amsmath, amssymb, bm}
\usepackage{hyperref}
\usepackage{multicol, wrapfig}

\graphicspath{{./figures/}}

\newcommand{\Photo}{}

\begin{document}
\vspace*{4cm}
\title{Bridging the Gap: Spectral Distortions meet Gravitational Waves}
\author{Thomas Kite}
\address{Jodrell Bank Centre for Astrophysics, School of Physics and Astronomy, The University of Manchester, Manchester, M13 9PL, U.K.}
\maketitle\abstracts{
This talk has the goal of introducing two upcoming exciting avenues of discovery in precision Cosmology: spectral distortions (SDs) and gravitational waves (GWs). The former signals offer a clear window into the state of the primordial plasma at times prior to recombination, and thus sheds light on small-scale primordial perturbations, dark matter decay and black hole formation to just mention a few scenarios. The latter signals, which are already offering new insights into the landscape of black hole and neutron star mergers, will reveal intricate dynamics of the early universe including primordial black holes, inflationary potentials, and even reheating dynamics. An elegant link is drawn between these two future observations since primordial GW backgrounds will source SDs, a coupling which offers unique insight to over six decades of GW frequencies. More importantly the SD visibility window bridges the gap between astrophysical high- and cosmological low-frequency measurements. This means SDs will not only complement other GW observations, but will be the sole probe of physical processes at certain scales.}
\vspace{-1.0cm}
{\small \section*{Disclaimer}
\vspace{-0.2cm}
This document has been written in such a way to mirror the talk presented at the 2022 Cosmology session of the 56th Rencontres de Moriond. In particular the focus is on qualitatively reviewing the main Physics and providing a more heuristic overview of the content. For a more complete and detailed discussion with mathematical details we point the reader to \cite{Kite:GW2SD} and references therein.}

\section{Spectral Distortions}
\vspace{-0.2cm}
For many decades the focus of precision Cosmology, at least in the early Universe, has been the power spectrum of Cosmic Microwave Background (CMB) temperature anisotropies \cite{WMAP:CMB_paper} \cite{Planck2018:overview}. The wealth of information contained in this signal has now been exquisitely well measured, allowing for the consolidation of a concordance model of Cosmology, \lcdm. 

It is a trivial yet important statement that there is only one CMB sky to observe, and as such it is imperative we tease as much information from it as possible. It is in this vein that we introduce spectral distortions (SDs) as one way of extending the enormous success of CMB observations for decades to come. The fundamental novelty in SD measurements is to step back from thinking of \textit{spatial} information content locked in angular distributions across the sky, and instead focus on the frequency distribution of the photon flux \footnote{Although future missions might detect SD anisotropies.}, which promises more information than a single-parameter blackbody.

A useful analogy for this is in expressing a function $f(x)$ as a Fourier series, where a continuous function thus is expressed as a countably infinite series of coefficients $f_i$. Only for a very special class of functions would you have a single non-zero coefficient. Similarly, the CMB photon flux spectrum across frequency is currently described using a single number: temperature. This is a natural and effective choice given that an interacting spectrum of photons will tend towards the thermal equilibrium described by a blackbody, but this only causes other \textit{CMB coefficients} to be small, not $0$. Standard \lcdm\ predicts \cite{chluba:which_SD_LCDM} primordial SD amplitudes $\sim 10^{-8}$.

\subsection{Primordial Origin of SDs}
Many processes in the universe can cause a deviation from a simple background blackbody (e.g. emission lines we observe from atomic transitions, up-scattered CMB photons upon colliding with hot gas). In this talk however we care about SDs from the primordial Universe which are usually decomposed in a few main types, each caused by energy injection in different eras.
\newline $\bullet$
The earliest energy injection into the photon bath ($z\geq 2\times10^6$) occurs in the very high temperature universe - a sufficiently energetic system to create photons (bremsstrahlung, double Compton emission) and redistribute their energies (Compton/Thompson scattering). This results in an unobservable shift to the blackbody spectrum to a higher temperature.
\newline $\bullet$
At later times ($5\times10^4\leq z\leq2\times10^6$) the mechanisms for photon production are no longer efficient even though photons can be redistributed. This results in a spectrum containing a $\mu$-distortion, equivalent to a chemical potential in an otherwise perfect blackbody.
\newline $\bullet$
The final era which roughly lasts all the way to recombination ($1100\leq z\leq5\times10^4$) is characterised by the lack of efficient energy redistribution mechanisms within the photon gas, leaving a $y$-distortion in the spectrum.

\subsection{Example: Dissipation of Acoustic Modes}
\begin{wrapfigure}{r}{0.5\linewidth}
    \vspace{-1.4cm}
    \centering
    \includegraphics[width=0.95\linewidth]{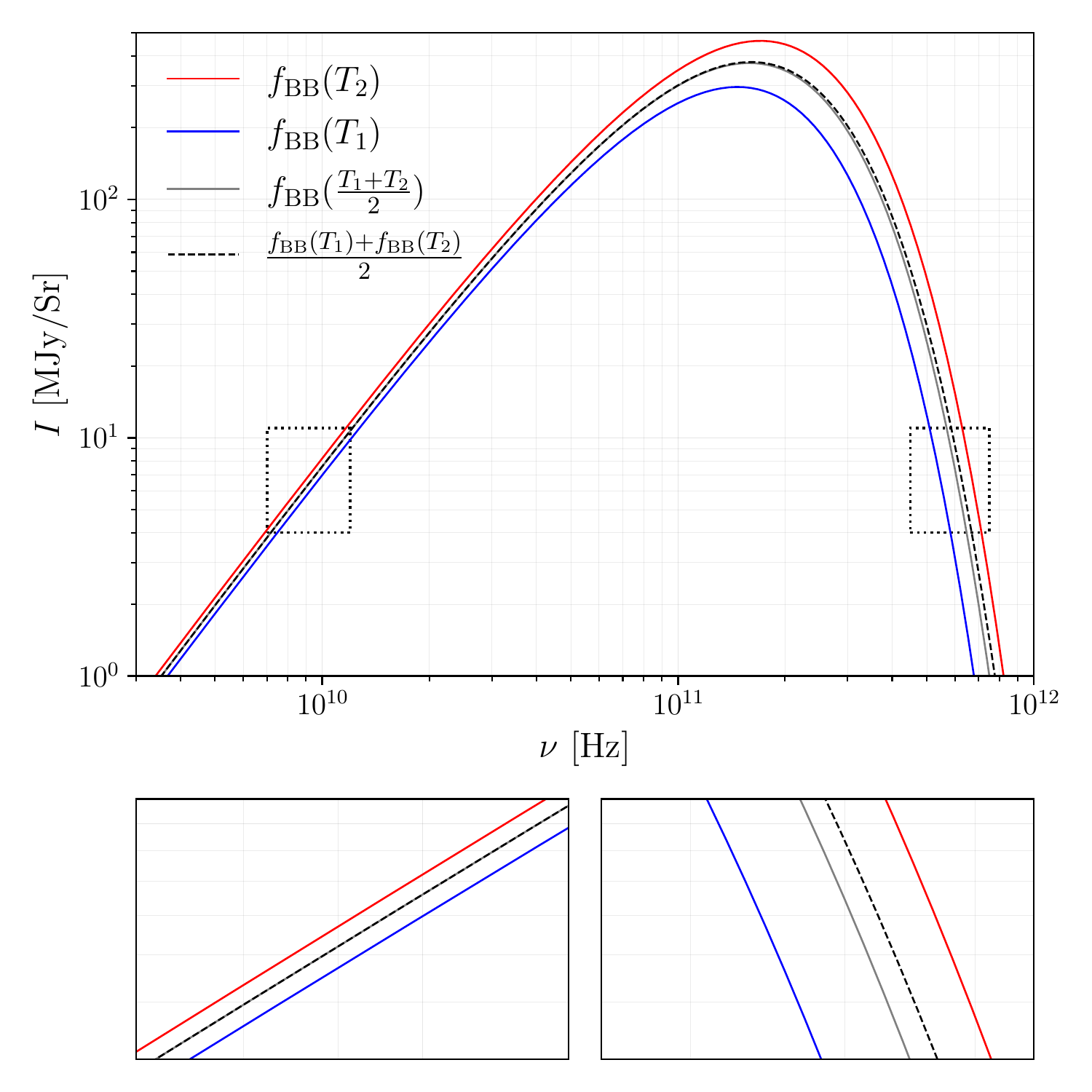}
    \caption{A plot illustrating how the mix of two blackbodies differs from the blackbody at the average temperature, rather there is an excess of high frequency photons. Figure adapted from \protect\cite{Chluba:new_physical_scales}}
    \label{fig:bb_mixing_figure}
\end{wrapfigure}
A primary source of SDs within \lcdm\ is from Silk damping of small-scale acoustic modes. Simply stated, small patches of differing temperature in the early Universe are sufficiently close to mix photons, thus erasing the temperature differences. This spatial isotropisation of the medium renders these patches invisible to CMB temperature anisotropies (hence the large $k$ damping in the CMB power spectrum), but a SD signal persists to be seen today. The key to understanding this process is that adding two blackbodies at $T_1$ and $T_2$ respectively will not exactly produce the expected $(T_1 + T_2)/2$ blackbody, but rather an excess photon tail remains at high frequency \cite{Chluba:new_physical_scales} as illustrated in Fig.~\ref{fig:bb_mixing_figure}. Those excess photons constitute an energy injection, which will remain in the spectrum as a SD.

\section{Gravitational Waves}
\vspace{-0.2cm}
The second upcoming cosmological probe this talk is concerned with is the gravitational wave background (GWB) \footnote{In contrast to single localised GWs, which are usually associated with individual astrophysical events.}. The key to the excitement in GW Cosmology is the simple fact that these tensorial waves interact extremely weakly with the intervening matter. Consider for example the existence of the \textit{CMB curtain}, which shields our view from the first $380$ thousand years of the Universe. This happens because the electromagnetic waves carrying the information interact too strongly with charged particles (which incidentally makes them easy to detect too). GWs have no such interaction, and it is very possible to see a GWB arising from inflationary dynamics or reheating mechanisms which travel to our detectors unhindered. This direct glimpse into the fundamental physics of the early-universe is the enticing proposal of GW Cosmology.

\subsection{Detection Prospects}

There are a diverse range of upcoming probes spanning over 20 orders of magnitude in frequency seeking to pin down the shape of the GWB \cite{Caprini:2018mtu}. On the lowest frequencies (largest scales) we have the cosmological scale probes with B-mode polarisation searches. Moving to higher frequencies (smaller scales) there is potential to glimpse GWs with astrophysical level probes like Pulsar Timing Arrays (PTA), astrometry and binary system resonances \cite{Blas:binary_resonance}. The highest of frequencies are covered by direct interferometry or even some solid state detectors. Each of these techniques are based on different physics, which links directly to the frequencies they're sensitive to.

An interesting \textit{gap} exists between the cosmological large-scale and astrophysical small-scale. Primordial waves in this window damp too early to leave notable impacts on the CMB polarisation. However, analogously to the previously discussed Silk damping, these waves excite a quadrupole moment in the photon multipole hierarchy, which ultimately causes an energy injection to the photon bath, and thus a SD signal.

\section{Bridging the Gap}
\begin{wrapfigure}{r}{0.5\linewidth}
    \vspace{-1.2cm}
    \centering
    \includegraphics[width=0.95\linewidth]{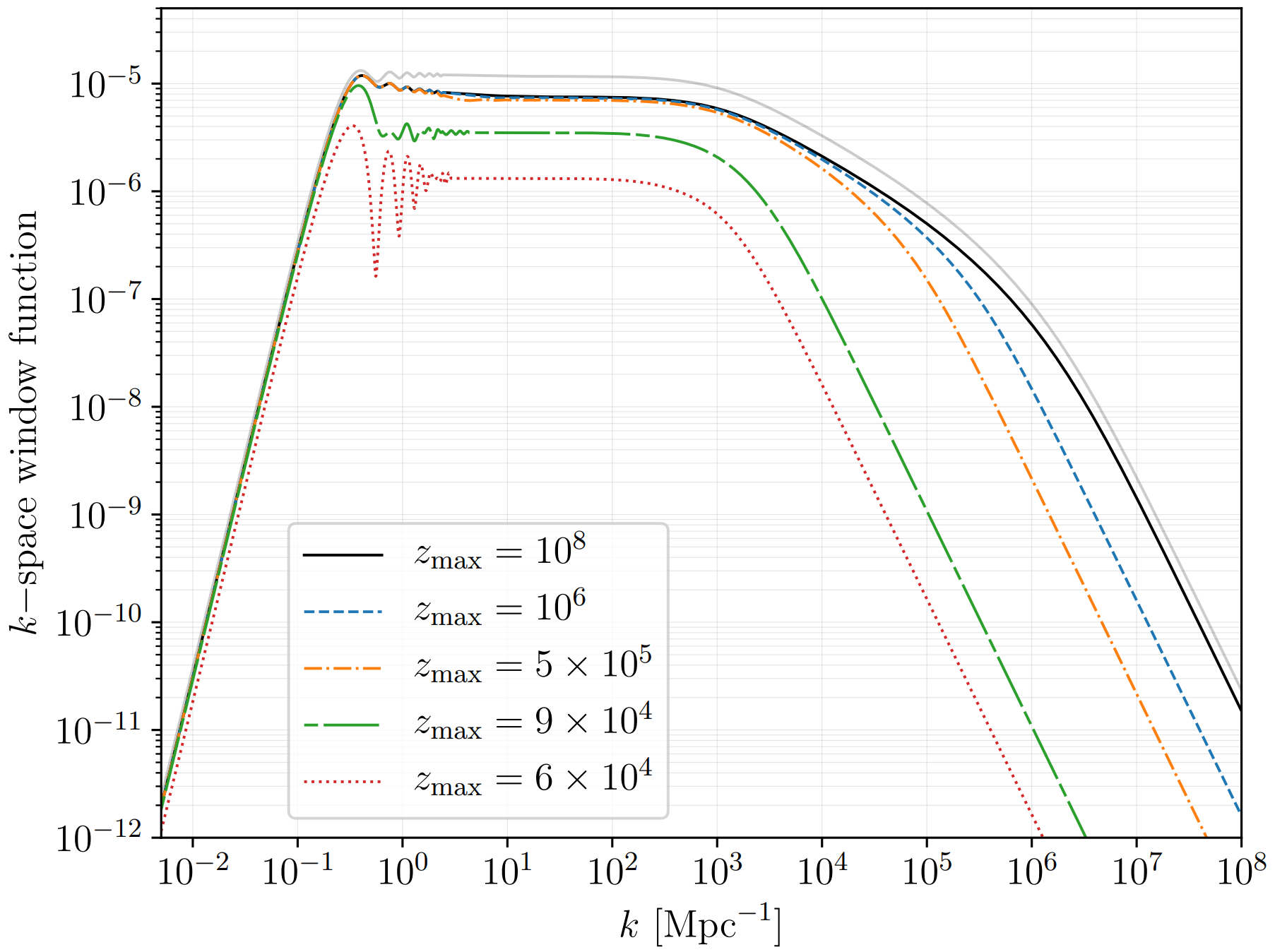}
    \caption{A plot showing the $\mu$-distortion window functions for GW power spectra. There is a plateau for $10^0 \lesssim k\,[{\rm Mpc}^{-1}] \lesssim 10^3$. Very late subhorizon GW injection reduces the total visibility by up to an order of magnitude, as seen by increasing $z_{\rm max}$. The gray line shows the window function if neutrino damping is neglected.}
    \label{fig:window_figure}
\end{wrapfigure}
%
In this section we will give some more mathematical detail about the calculation of SD arising from a GWB while still leaving main technical details to \cite{Kite:GW2SD} \cite{Chluba:tensor_paper}. The bottom line is that for a given cosmology a lot of the calculation can be precomputed, leaving only a single integral over wavenumber $k$ of the tensor power spectrum $\mathcal{P}_{T}(k)$ multiplied by a SD window function $W_{\mu}^{z_{\rm max}}(k)$:
\begin{equation}
    \langle\mu_{\rm GW}\rangle=\int_{0}^{\infty}{\rm d}\ln k\,\,\mathcal{P}_{T}(k) W_{\mu}^{z_{\rm max}}(k)\,. \label{eq:mu_GW}
\end{equation}
The window function is calculated by combining three main terms:
\begin{equation}
    \mathcal{W}_{\mu} = \left(1.401 \mathcal{J}_{\mu}\right) \left(\frac{8 H^{2}}{45 \dot{\tau}} h^{\prime 2}\right) \left(\mathcal{T}_{\Theta} e^{-\Gamma_{\gamma}^{*} \eta}\right)\,, \label{eq:window_primative}
\end{equation}
which from left to right correspond to: firstly a \textit{SD branching ratio} which dictates the resulting $\mu$-distortion amplitude from some energy injection, secondly a term corresponding to the energy content of a GW, and thirdly the term expressing the coupling between GWs and photons. Integrating this over the relevant redshifts while keeping in mind an upper redshift cutoff for any GWB created subhorizon,
\begin{equation}
    W_{\mu}^{z_{\rm max}}(k) = \int_{0}^{z_{\rm max}}\mathrm{d}z\,\, \mathcal{W}_{\mu}(k,z)\,, \label{eq:window_integral}
\end{equation}
gives the full SD window function. These window functions are shown in Fig.~\ref{fig:window_figure}.

To facilitate this type of calculation we provide the precomputed window functions for standard \lcdm\ using best fit \textit{Planck18} cosmological parameters \cite{Planck2018:parameters} via a simple python tool {\tt GW2SD} (\url{https://github.com/CMBSPEC/GW2SD.git}).

\begin{figure}[h]
    \centering
    \includegraphics[width=\linewidth]{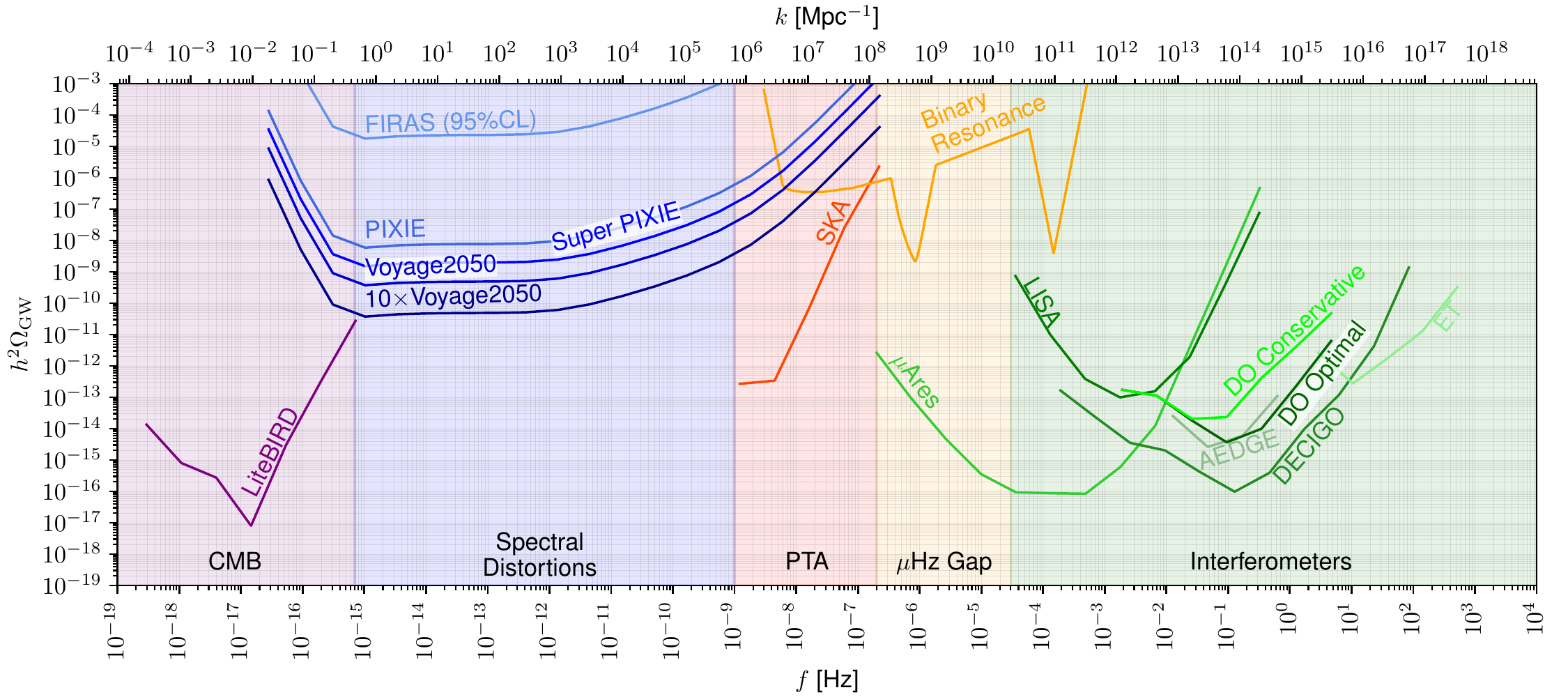}
    \caption{A plot showing the different probes of the GWB in their respective frequency bands. From left to right (low to high frequency, large to small-scales) there are CMB B-modes, CMB spectral distortions, pulsar timing arrays, binary resonance searches and direct detection interferometry. The SD window bridges the gap between the cosmological large-scale probes and the astrophysical small-scale. Sensitivity curves are taken from \protect\cite{Kite:GW2SD} \protect\cite{Caprini:2018mtu} \protect\cite{Blas:mu_gap}.}
    \label{fig:GWB_figure}
\end{figure}

Mapping these SD window functions into sensitivity curves reveal how SDs complement other probes in the GWB landscape, as shown in Fig.~\ref{fig:GWB_figure}. As previously discussed SDs are able to bridge the gap between high- and low-frequency probes. This means that, despite their overall lower sensitivity (GWs and photons only interact feebly) SDs will provide a unique or complementary probe of new physics on certain energy scales. Furthermore, with a wide sensitivity curve it is possible to integrate sufficient power over many frequency bins to produce a measurable $\mu$-distortion signal from a GWB that would otherwise be too weak to detect. To illustrate these points we investigate a few models of interest in \cite{Kite:GW2SD}, and show which regions of parameter space can be constrained with either future data or even using the three decades old data available from {\it COBE/FIRAS} \cite{Fixsen:original_T0_paper}.
\vspace{-0.5cm}
{\small
\section*{References}
\bibliography{moriond}
}

\end{document}